\documentclass[10pt]{article}
\pdfoutput=1
\usepackage{jheppub}
\usepackage{amsmath,amssymb,amsfonts,mathbbol,graphicx,slashed,color,amsthm, mathtools, upgreek, enumerate, tensor}
\usepackage{subcaption}
\usepackage{setspace}
\usepackage[export]{adjustbox}
\usepackage{arydshln}
\usepackage[dvipsnames]{xcolor}
\usepackage{physics}
\usepackage{comment}
\graphicspath{{pics/}}

\allowdisplaybreaks

\colorlet{darkblue}{blue!70!black}

\colorlet{darkgreen}{green!50!black}
\colorlet{darkred}{red!50!black}

\def\bea{\begin{eqnarray}}
\def\eea{\end{eqnarray}}
\def\be{\begin{equation}}
\def\ee{\end{equation}}

\title{
Black hole to cosmic horizon microstates in string/M theory: timelike boundaries and internal averaging 
}

\author[a]{Eva Silverstein}
  \affiliation[a]{Stanford Institute for Theoretical Physics, 382 Via Pueblo, Stanford, CA 94305}
 \vspace{5mm}

\vspace{1cm}

\emailAdd{evas@stanford.edu}

\abstract{

In this note, we resolve an apparent obstacle to string/M theory realizations of dS observer patch holography, finding a new role for averaging in quantum gravity. 
The solvable $T\bar T(+\Lambda_2)$ deformation recently provided a detailed microstate count of the $dS_3$ cosmic horizon,  reproducing the refined Gibbons-Hawking entropy computed by Anninos et al along with the correct radial bulk geometry.  On the gravity side, the deformation brings in the boundary to just outside a black hole horizon, where it is indistinguishable from the dS cosmic horizon, enabling a continuous passage to a bounded patch of dS.  In string/M theory, the relationship between AdS/CFT and dS involves uplifts that change the internal topology, e.g. replacing an internal sphere $\mathbb{S}$ with an internal hyperbolic space $\mathbb{H}$ (and incorporating varying warp and conformal factors).  We connect these two approaches, noting that the differences in the extra dimensions between AdS black hole and dS solutions are washed out by internal averaging in the presence of a timelike boundary skirting the horizon.  
This helps to motivate a detailed investigation into the possibility of such timelike boundaries in (A)dS solutions of string/M theory, and we take initial steps toward suitable generalizations of Liouville walls as one approach.

}

\begin{document}

\maketitle
\parskip=10pt

\section{Introduction}

In quantum gravity, methods ranging from low-dimensional toy models to full-fledged string/M theory provide insight in very different regimes.  In this note, we explain a basic consistency check between them in the case of de Sitter (dS) static patch holography.    

The universal contribution to the one-loop corrected three-dimensional de Sitter ($dS_3$) horizon entropy \cite{Anninos:2020hfj} has been computed independently by an explicit, solvable deformation of 2d holographic CFTs  \cite{Coleman:2021nor}. In addition to producing the correct state count, this $2d$ non-gravitational, non-local boundary theory captures the emergent radial geometry of a bounded region of the static patch by matching its quasilocal energy as a function of the boundary position. This method replaces BPS control used for special black hole microstate counts \cite{Strominger:1996sh} with integrability of solvable deformations, pioneered by works such as  \cite{Zamolodchikov:2004ce}\cite{Smirnov:2016lqw}\cite{Dubovsky:2012wk}\cite{Cavaglia:2016oda}.\footnote{The groundwork for this result was laid in e.g. \cite{Gorbenko:2018oov}\cite{Shyam:2021ciy}, and independent analyses such as \cite{Svesko:2022txo, Banihashemi:2022jys, Banihashemi:2022htw, Pasquarella:2022ibb} recently recovered similar and/or supporting conclusions; additional insights into the physics appeared in \cite{Coleman:2022lii}, a reformulation in terms of an auxiliary 2d gravity path integral will appear \cite{to-appear-Gonzalo-kernel},  and a small sample of related work on dS entropy includes \cite{Banks:2006rx, Banks:2016taq, Anninos:2011af, Dong:2018cuv, Anninos:2020geh, Anninos:2021eit, Anninos:2021ihe, Shaghoulian:2021cef, Anninos:2022ujl, Chandrasekaran:2022cip}.}  In this note, we show how an apparent obstacle to generalizing this method to string/M theory is neatly averted, and initiate a research program to construct the required UV complete timelike boundaries.    

The model \cite{Coleman:2021nor} explicitly captures the aforementioned properties -- the finite state count and radial geometry -- but by itself does not produce a theory with non-gravitational local bulk matter. It is known how to address this in perturbation theory about large N \cite{Hartman:2018tkw}.  Moreover, one may be able address this beyond large N in several ways.  For example, the space of solvable models may extend to include bulk gauge fields and/or bulk conformal fields -- a topic of current investigation \cite{WIP-bulk-matter}.   Lower dimensional examples building from \cite{Gross:2019ach} have the advantage of less ambiguity from irrelevant deformations, potentially making contact with approaches such as \cite{Susskind:2021dfc, Susskind:2021esx}.  In any case, with string/M theory providing higher-dimensional UV complete quantum gravity -- a theory which is massively dominated by positive potential energy contributions to the $4d$ effective potential \cite{Flauger:2022hie} -- it is important to assess this (or any other) approach to cosmological holography in that framework. 

\begin{figure}[t!]
  \centering
  \includegraphics[width=0.95\linewidth]{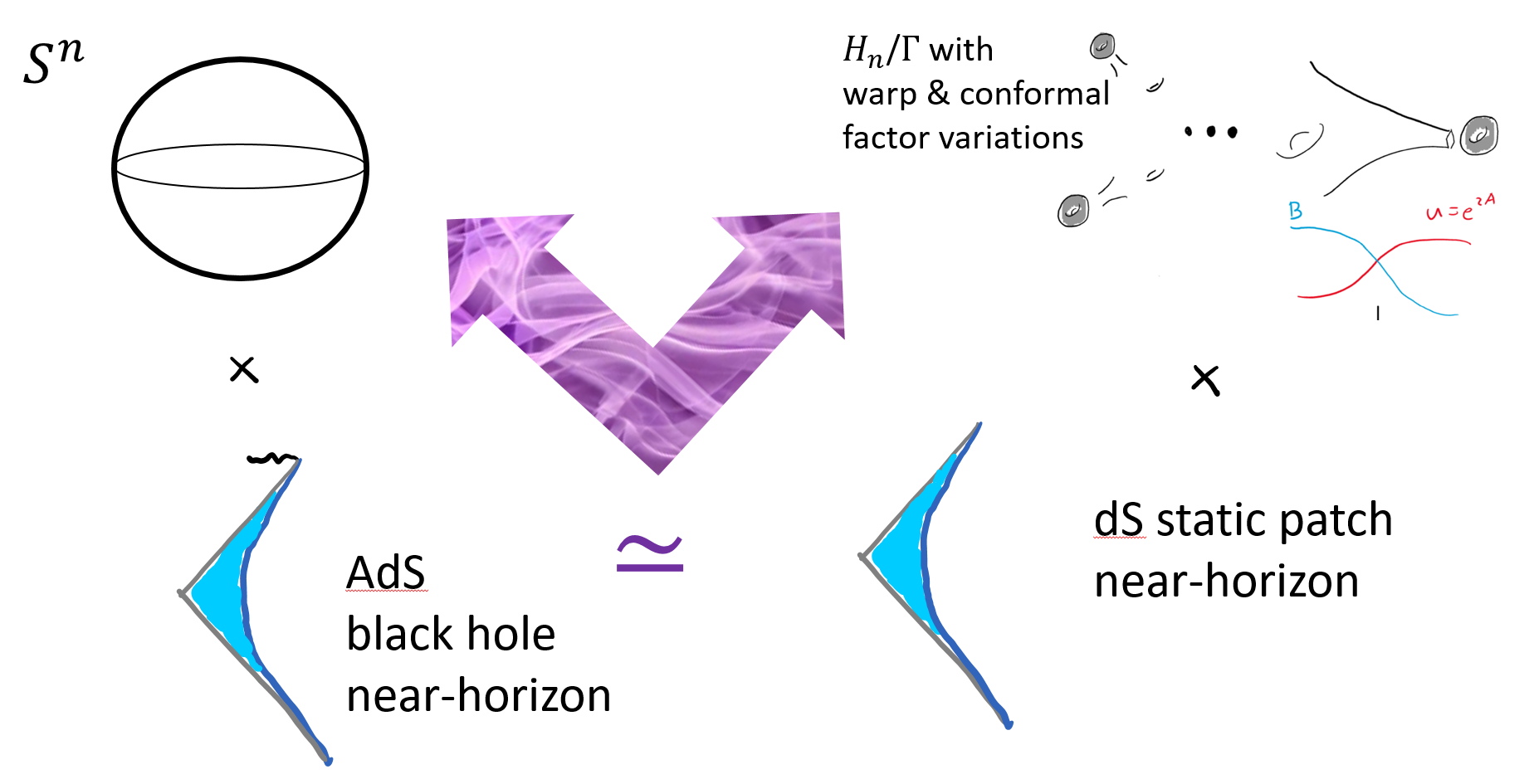}
  \caption{ \small \sffamily 
  Although their near-horizon geometries are indistinguishable in the external dimensions, the semiclassical internal solutions for AdS black holes and dS are very different \cite{Flauger:2022hie}, as exemplified by the uplift from an internal sphere to internal hyperbolic space \cite{DeLuca:2021pej}.  Nonetheless, this internal distinction also disappears at the near-horizon matching point in the deformation \cite{Coleman:2021nor} between AdS black holes and dS. 
 % obtained by bringing in the AdS boundary to the black hole, and moving it out in the dS static patch.  
 This is because the near-horizon boundary entails a nearly diverging boundary temperature, compelling a thermal mixing among the internal configurations connected by string/M theoretic topology changing transitions.  Similarly, at fixed boundary quasilocal energy the system is not in an eigenstate of internal topology}
  \label{fig:internal-mixing}
\end{figure}

The relevant $T \bar T$-type deformation \cite{Coleman:2021nor} formulates the dual of bringing in the $AdS_3$ boundary to a point where $AdS_3$ and $dS_3$ become indistinguishable, then bringing the boundary out in a way that captures the $dS_3$ static patch geometry with $T\bar T+\Lambda_2$.
For the entropically dominant energy level, a key step step in the derivation \cite{Coleman:2021nor} is the matching point between these two processes, where the theory transitions from the dual of an AdS black hole horizon to the dual of the dS cosmic horizon.   These are indistinguishable when the patch boundary skirts the horizon (see figure \ref{fig:internal-mixing}).\footnote{Note that for microstates, the region beyond the horizon is generically non-geometrical, singular at the horizon \cite{Czech:2012be}.} But one might worry that this approach could never generalize to string/M theory, in which there are significant internal differences among AdS and metastable dS solutions.\footnote{See e.g. section 2 of \cite{Flauger:2022hie} and references therein for a recent summary of the relevant string/M theory compactifications.}  
In the present work, we will point out that this step does generalize to string/M theory, conditioned on the existence of the requisite patch boundaries in the theory.  The reason is conceptually simple: this transition point corresponds in gravity-side variables to a bounding wall at the (stretched) horizon, where its acceleration relative to an inertial observer in the bulk reaches the string or Planck scale, compelling an effective thermal mixing among all physically connected internal configurations (including topology changing processes).  Interesting recent work on a 2d analogue of this, with a study of internal dilatonic degrees of freedom involved in (A)dS transitions appeared in \cite{Pasquarella:2022ibb}.  
% In the canonical ensemble obtained from an appropriate mixed state of microstates, this effect would arise from thermal averaging
This introduces a new role for thermal averaging in quantum gravity. It helps  motivate a research program to assess the UV completion of fixed boundary conditions and quasilocal notions of energy in general relativity, the latter itself a subject of much recent activity (see e.g. \cite{Marolf:2012dr, Anderson:2006lqb, Andrade:2015gja, An:2021fcq, Witten:2018lgb, Marolf:2022jra, Marolf:2022ntb}).  

The problem of classifying timelike boundaries in quantum gravity beyond asymptotically AdS solutions is important in any case.  The claim is often heard that a positive cosmological constant precludes timelike boundaries, but this is a highly nontrivial and unproven assertion.  (What is true is that global de Sitter spacetime has no timelike boundary, a different statement.)  If the solvable models --with their explicit match of quasilocal energies and corrected entropy -- are any guide, such boundaries can play an important role in generalizing holography to cosmology.  In any case, it is a technical question to assess the existence or not of such bounded string backgrounds.           

The organization of the paper is as follows.  In \S\ref{sec:internal-equilibration} we explain the conditional generalization to string/M theory of the deformation between AdS black holes and the dS cosmic horizon.  In \S\ref{sec:string-M-boundaries} we set up the problem of UV complete timelike boundaries, and initiate one approach to it via generalizations of Liouville theory \cite{Harlow:2011ny, Teschner:2001rv,Ginsparg:1993is, Seiberg:1990eb} in some examples.  We finish with a discussion of future directions, including a sketch of the lift of this approach to M theory via Matrix (String) Theory \cite{Banks:1996vh, Dijkgraaf:1997vv} related to work in progress \cite{WITP-M-HW-walls}.

\section{The transition from an AdS black hole to dS:  Internal averaging}\label{sec:internal-equilibration} 

Let us entertain the possibility that string/M theory allows for a boundary on a cylinder surrounding an AdS black hole and also allows for a bounding cylinder in the dS static patch, generalizing the gravity side of \cite{Coleman:2021nor} to $d=3$ or 4 external dimensions \cite{Dong:2010pm, DeLuca:2021pej}.  In this section we will start from this and one other general assumption:

\begin{enumerate}\label{assumptions}
    \item {\it Bounding walls exist, with boundary conditions compatible with fluctuations in the internal space.} Their existence is an open question, amenable to technical analysis, including the approach discussed in \S\ref{sec:string-M-boundaries} and \S\ref{sec:discussion}.    
      
    \item {\it The string/M theory configuration space is connected, implying the existence of physically accessible topology change between different compactifications such as $\mathbb{S}^7$ and $\mathbb{H}_7/\Gamma$.}  Significant precedents for this statement exist \cite{Aspinwall:1993nu, Witten:1993yc, Distler:1995bc, Strominger:1995cz, Kachru:1997rs, Adams:2005rb}, including connections between positive, flat, and negatively curved spaces \cite{Adams:2005rb} and among dimensionalities (values of the effective central charge) \cite{Hellerman:2006hf}.   
\end{enumerate}

Specifically, we suppose there may exist a wall at fixed radius $r=r_c$ in the $d+1$-dimensional AdS black hole metric
\be\label{eq:bhads}
ds_{AdS-BH_{d+1}}^2= \frac{dr^2}{f(r)^2}-f(r)^2 dt^2+r^2 d\Omega_{d-1}^2\,,
\ee
with
\be\label{eq:fdef}
f(r)^2 = 1+ \,\frac{r^2}{l^2}-\frac{16 \pi\, G}{(d-1) \text{Vol}(S^{d-1})}\,\frac{M}{r^{d-2}}\;, 
\ee
and also there may be a UV complete wall at fixed radius $r=r_c$ in the dS static patch
\be\label{eq:ds-metric}
ds_{dS_{d+1}}^2 = \frac{dr^2}{1-\frac{r^2}{\ell^2}}-(1-\frac{r^2}{\ell^2}) dt^2+r^2 d\Omega_{d-1}^2\,,
\ee
We are particularly interested in the case where the wall skirts the horizon in these backgrounds.

A priori, the boundary conditions at such walls may take a variety of forms.  A key distinction is whether or not one integrates over the boundary values of the fields.\footnote{We thank G. B. De Luca, V. Shyam, G. Torroba and S. Yang for many discussions of this at the bottom up level and ongoing research on the spectrum of possibilities for bounding walls \cite{WITP-M-HW-walls}.}  In the Horava-Witten construction \cite{Horava:1996ma} one does such an integral, forcing the Brown-York stress energy to cancel against stress-energy from explicit boundary degrees of freedom.
In contrast, a Dirichlet condition on the metric freezes it at the boundary and is consistent with the bounded patch carrying non-vanishing net stress-energy; this case arises in the holographic interpretation of $T\bar T+\Lambda_2$ \cite{McGough:2016lol, Kraus:2018xrn, Gorbenko:2018oov, Lewkowycz:2019xse, Coleman:2021nor}.  A similar distinction arises for other bulk fields including gauge fields.  The Dirichlet case would be most analogous to AdS/CFT holography, but other types of boundaries may also be interesting.  For our purpose in this section, we require a boundary that allows the internal configuration to fluctuate.  We will return to this point in the context of the approach to constructions that we take below in \S\ref{sec:OLT-effective-bc}.          

In the gravity-side variables, the deformation between AdS/CFT and dS/deformed-CFT  in \cite{Coleman:2021nor} involves a first step of bringing in the AdS boundary in to the horizon ($r_c\to r_{horizon}$) of a black hole at the Hawking-Page transition, with horizon radius equal to the AdS curvature radius $\ell$.  
This near horizon patch is identical for an AdS black hole and dS with the same horizon radius $\ell$.  Starting from this configuration, the second step is to bring out the boundary to obtain the cosmic horizon patch of dS.  (In the language of the boundary theory, the first step is a pure $T\bar T$ deformation in the boundary language, matching to a $T\bar T+\Lambda_2$ deformation for the second step \cite{Coleman:2021nor}.)   
Given our assumption (1) that bounding walls exist in string/M theory,  this gravity-side procedure, including the manifest indistinguishability at the matching point where the boundary skirts the horizon, generalizes immediately to its $d+1\ge 3$ external dimensions.  

However, at first glance this procedure may nonetheless seem impossible to embed into string/M theory, because of the obvious {\it internal} differences between compactifications to dS versus AdS.  For example, the $AdS_4\times \mathbb{S}^7$ Freund-Rubin solutions of M theory with 7-form flux uplifts to a warped product of $dS_4$ with a hyperbolic internal space $\mathbb{H}_7/\Gamma$ (conformally deformed and Anderson-Dehn filled \cite{Anderson:2003un}, tuned so that the net curvature competes with Casimir energy and 7-form flux \cite{DeLuca:2021pej}).  The topology change required to interpolate between the $\mathbb{S}^7$ and the $\mathbb{H}_7/\Gamma$ -- if it exists similarly to \cite{Adams:2005rb} --  would be singular in the low energy supergravity theory; it introduces high barriers between them in the string/M configuration space.

Given such an interpolation, the indistinguishability does extend to the internal dimensions, because of the fact that the matching point is at a diverging temperature.  The point is:
\begin{quote}
    {\it The hot boundary skirting the horizon compels a thermal mixture, averaging among all physically connected internal spaces consistent with this horizon.} 
\end{quote}
Thus, given the pair of assumptions  (\ref{assumptions}):  the existence of bounding walls and of topology changing processes, we can indeed interpolate between AdS/CFT and dS/deformed-CFT in the full higher-dimensional string/M theory.  This argument is phrased in the canonical ensemble, but one may transform to the microcanonical ensemble.  At fixed energy, in the near horizon region there is no reason that fixed-energy states correspond to a definite topology, leading to a similar effect that the internal differences are washed out.  We note that in our setup, the mixing happens everywhere in the bounded near-horizon patch, rather than realizing a domain wall solution.\footnote{Such solutions -- which would generically be time-dependent -- would be interesting to study in their own right.}

Related to this, a similar question arises in formulating the complementary region within the static patch, the pole patch \cite{Coleman:2021nor}.  There, the matching between $T\bar T$ and $T\bar T +\Lambda_2$ occurs at a point in the trajectory described on the gravity side by a small cylinder in vacuum AdS (near the center) or the dS static patch (near the pole).  Again, this small cylinder in vacuum AdS or dS is equivalent in the external dimensions, whereas the internal dimensions are very different.  In this case, we expect the resolution is similar to the above: at the matching point with the small cylinder-bounded patch, energy eigenstates will not generally coincide with fixed topology, given the connected configuration space, washing out the internal difference.

% {\color{blue} {\bf Clarify ensemble, precise boundary condition choice(s) needed, etc}}

The assumptions (\ref{assumptions}) require further research to assess.  Even from the bottom up (GR coupled to effective fields), the status of various boundary conditions is under active investigation  (see e.g. \cite{Marolf:2012dr, Anderson:2006lqb, Andrade:2015gja, An:2021fcq, Witten:2020ert, Marolf:2022jra, Marolf:2022ntb}).  Interestingly, some of these considerations are UV sensitive.  In the remainder of this paper, we exhibit some new examples of boundaries, taking the approach of generalizing Liouville theory for this purpose. We leave a systematic study of all cases to future work.  

% *****************

% *****************

% the effectively infinite temperature washes out the distinction between AdS and dS both externally (where the horizon is Rindler) and internally (the hot boundary skirting the horizon compels a thermal mixture, averaging among all internal spaces consistent with this horizon). One can then continue the deformation to bring the boundary out to recover the dS static patch, along the way perhaps settling the question of the relevant algebra (type I or II).  In addition to the existential question of such boundaries and quasilocal energies in GR and string/M theory  \textbf{C,D}, the high-temperature internal mixing requires topology change, bringing in modern studies of such processes.

% The model \cite{Coleman:2021nor} captures the $3d$ gravity sector.  The gravitational dual of the two-part $T\bar T$ and $T\bar T+\Lambda_2$ trajectory   boundary is brought in to the AdS black hole horizon. 

% The purpose of the present paper is twofold

% Much research on boundary conditions in general relativity 

\section{Toward timelike boundaries in string/M  theory}\label{sec:string-M-boundaries}

We are ultimately interested in placing a timelike boundary with cylindrical geometry in an $AdS_{d+1}$ black hole and in the $dS_{d+1}$ static patch (in a system of some higher total dimensionality $D$).
In this section, we will first formulate this problem in some generality and discuss various issues.   Then we will study in more detail some simpler examples admitting chiral currents, focusing on the question of total reflection at the wall. We will comment on generalizations in the discussion section.  

\subsection{Generalities}

% {\bf (Alternatively, start with simple case and later generalize) }

Including all dimensions, these spaces take the form
\begin{equation}\label{eq:generalspace}
    ds^2 = e^{2 A(y)}\left\{d\chi^2+ f_1(\chi)(-dt_\parallel^2+f_2(\chi){(d\vec x_\parallel)^2}) \right\}+g_{mn}(y)dy^m dy^n
\end{equation}
with the indices $m,n$ running over the internal $D-(d+1)$ dimensions.  In these coordinates, the bounding wall of interest would appear at $\chi=\chi_c$ and the $\parallel$ notation indicates direction along the wall.  In addition, there is flux, and other ingredients depending on the example \cite{Flauger:2022hie}.  

There is much recent and ongoing research on the boundary value problem in general relativity (see e.g. \cite{Marolf:2012dr, Anderson:2006lqb, Andrade:2015gja, An:2021fcq, Witten:2020ert, Marolf:2022jra, Marolf:2022ntb}).  In the present note, we explore potential methods for constructing string/M theoretic bounding walls.  We will focus first on string-theoretic examples, and comment on lifts to M theory in the discussion section.  These bottom up and top down analyses overlap at low energies, but some of the issues appear UV sensitive (e.g. involving short modes at the boundary \cite{Anderson:2006lqb, Witten:2020ert}).    

In this note, our main approach technically will be to generalize Liouville walls\footnote{For reviews and solutions of standard Liouville theory see e.g. \cite{Seiberg:1990eb}\cite{Ginsparg:1993is}\cite{Harlow:2011ny} and \cite{Dorn:1994xn, Zamolodchikov:1995aa}.} to higher-dimensional target spaces.    
In classic Liouville and Sin-Liouville theory, the worldsheet action at the semiclassical level contain a term $\sim \int d^2\sigma \hat O e^{\kappa\chi}$, with $\kappa$ chosen so that the operator is marginal with respect to the flat bulk target spacetime theory.  The full path integral generates a 2d CFT in those cases, with structure constants for the pure Liouville theory proposed in \cite{Dorn:1994xn, Zamolodchikov:1995aa}.  The interaction turns off in the bulk where $\chi$ becomes large and negative, with incoming and outgoing wave functions defined in this region.  They totally reflect at the wall, with a phase computed in \cite{Dorn:1994xn, Zamolodchikov:1995aa}.   

Let us
consider more generally a worldsheet theory whose semiclassical action is of the form
\begin{equation}\label{eq:ws-def-gen}
    S_{ws}=S_{ws}^{(0)}+\int_{\Sigma}[{\hat O}_\Delta\Phi]_r
\end{equation}
where $\Delta>2$ is the dimension of $\hat O$ in the unwalled CFT defined by $S_{ws}^{(0)}$, and $\Phi[\chi]$ is a function of the target spacetime embedding coordinate $\chi$ transverse to the wall, which solves the leading large-radius equations of motion for a field of mass corresponding to $\Delta$, $m^2 =2\Delta/\alpha'$.  The added term in \eqref{eq:ws-def-gen} is similar to a massive vertex operator (renormalized as in \cite{Polchinski:1998rq, Polchinski:1998rr}), but is non-normalizable due to the dependence only on $\chi$.    
In the simplest cases treated below, we can construct $\hat O$ out of certain chiral isometry currents in the unwalled theory, and $\Phi[\chi]$ is in a non-normalizable representation of the full current algebra, with $m^2$ related to the Casimir of the representation.  Our main concern in this work is to exclude the possibility that the string reaches $\chi\to\infty$ in some configuration where $\hat O$ vanishes.  

Looking ahead to future work, in the de Sitter target space case, the worldsheet description is more subtle, even in power-law stabilized string theory examples such as \cite{Maloney:2002rr, Dong:2010pm} since they involve Ramond-Ramond sector fields and a Fischler-Susskind type mechanism \cite{Fischler:1986ci, Fischler:1986tb} for worldsheet conformal invariance which would be interesting to incorporate.  A potentially simpler option is to build up from string theory to M theory along the lines we discuss in \S\ref{sec:discussion} and address walls in the M2-brane version of AdS/CFT \cite{Maldacena:1997re} and its uplift \cite{DeLuca:2021pej} to $dS_4$.    

\subsection{Walls in string theory}

We will start with flat Liouville-type walls in a flat target space, which may be a prototype for the more general case of weakly curved backgrounds with a string-scale wall.  We then generalize to vacuum $AdS_3$ and make some comments about the more general case.   In these examples, we will set up candidate worldsheet theories for walled backgrounds, focusing on the necessary condition that the string totally reflect at the putative wall.  

\subsubsection{Flat walls in flat target spacetime}\label{sec:flat-flat}

In a flat target space metric 
\begin{equation}
    ds^2 = -{dX^0}^2 + \sum_{i=1}^{D-1} {dX^i}^2
\end{equation}
we can put a candidate flat wall transverse to the $X^{D-1}$ direction via a worldsheet theory such as
\begin{eqnarray}\label{eq:ws-def-flat}  
    % S_{ws} &= S_{ws}^{(0)}-\lambda_W\int_{\Sigma}[
    % (
    %{J^0_L}J^0_R-
    % \sum_{i=1}^{D-1}{J^i_L}{J^i_R})^2 e^{\kappa X^{D-1}}]_r 
        S_{ws} &=& S_{ws}^{(0)}-\lambda_W\int_{\Sigma}[\{(-{J^0_L}^2+\sum_{i=1}^{D-2}{J^i_L}^2)(-{J^0_R}^2+\sum_{i=1}^{D-2}{J^i_R}^2)\}^2 e^{\kappa X_{D-1}}]_r \\ \nonumber
        &=& S_{ws}^{(0)}-\lambda_W\int_{\Sigma} [ \{(\partial_-X_\parallel)^2 (\partial_+X_\parallel)^2\}^2 e^{\kappa X_{D-1}}]_r
\end{eqnarray}
where $J^M_L=\partial_- X^M$ and $J^M_R=\partial_+ X^M$.  This is at the bosonic level, shortly we will detail the fermionic contributions for the case of worldsheet-supersymmetric theories. 
The subscript $\parallel$ refers to the directions transverse to $X^{D-1}$, parallel to the wall; the background preserves Lorentz invariance in those directions.   The value of $\kappa$ is fixed to render the added integrand marginal, i.e. dimension (1,1) in the unwalled $S^{(0)}$ theory:  
\begin{equation}\label{eq:kappa-flat}
    \kappa = \sqrt{\frac{2\Delta -4}{\alpha'}}=\frac{2 \sqrt{3}}{\sqrt{\alpha'}}
\end{equation}
In the fermionic string theories, the constraints of worldsheet supergravity require the vanishing of the simple pole in the OPE of the supercurrent $G=\psi^M \eta_{MN}\partial X^N$ with the $\lambda_W$ term in the Lagrangian (and similarly for the right movers).  The worldsheet-supersymmetric version of \eqref{eq:ws-def-flat} may be obtained by shifting $\partial X^M \to \partial X^M - \frac{1}{4} \psi^M \kappa \psi^{D-1}$ there.   
%Rotating to Euclidean signature on the worldsheet, this gives a path integral measure
% \begin{equation}\label{eq:lag-euclidean}
%     % \exp(-\frac{1}{\alpha'}S_{ws, E}) = \exp(-\frac{1}{\alpha'}\int d^2\sigma_E \left\{\dot X^2+(X^\prime)^2 +\lambda_W [\dot X_\perp^2+(X_\perp^\prime)^2]^2e^{\kappa X^{D-1}}  \right\}) 
%     \exp(-\frac{1}{\alpha'}S_{ws, E}) = \exp(-\frac{1}{\alpha'}\int d^2\sigma_E \left\{\dot X^2+(X^\prime)^2 +\lambda_W [\sum_{i=1}^{D-1}(({\dot X}^i)^2+({X}^\prime)^{2})]^2e^{\kappa X^{D-1}}  \right\}) 
% \end{equation}
% where the subscript $\perp$ refers to the directions $\mu=0,1, \dots, D-2$ transverse to $X_{D-1}$.   
% Computing the addition to the Hamiltonian density of this worldsheet action, using $\frac{\partial J^a_L}{\partial\dot X^b}=\delta^a_b=\frac{\partial J^a_R}{\partial\dot X^b}$, we obtain
% \begin{equation}\label{eq:Delta-H}
%  \Delta {\cal H} =\lambda_W e^{\kappa X_{D-1}}\left\{ ({J^0_L}^2+\sum_{i=1}^{D-1}{J^i_L}^2)({J^0_R}^2+\sum_{i=1}^{D-1}{J^i_R}^2) + (\sum_{M=0}^{D-1} J^M_L J^M_R)[ ({J^0_L}^2+\sum_{i=1}^{D-1}{J^i_L}^2)+({J^0_R}^2+\sum_{i=1}^{D-1}{J^i_R}^2)]\right\}  
% \end{equation}
The equation \eqref{eq:ws-def-flat} is written in conformal gauge on the worldsheet.  The covariant version for a general worldsheet metric $h$ is
\begin{equation}\label{eq:ws-def-flat-covar}
    S^{0}_{ws}-\frac{1}{4}\lambda_W\int d^2\sigma \sqrt{-h}\{(J_\parallel^2)_{\alpha\beta}(J_\parallel^2)_{\gamma\delta}(h^{\alpha\gamma}h^{\beta\delta}-\frac{1}{2}h^{\alpha\beta}h^{\gamma\delta})\}^2 e^{\kappa X^{D-1}} 
\end{equation}
where $(J_{\parallel}^2)_{\alpha\beta}=-J^0_
\alpha J^0_\beta + \sum_{i=1}^{D-2} J^i_\alpha J^i_\beta$.  

Bulk vertex operators in this system are defined as in the standard superstring in the region $X^{D-1} \to -\infty$ where the effects of the candidate wall term decay exponentially.  In this setup, we can exclude the possibility of a finite-energy string propagating to $X_{D-1}\to\infty$. 
In order to propagate to large $X^{D-1}$, we would need the non-negative coefficient $\hat O_8$ of the Liouville factor in \eqref{eq:ws-def-flat} to vanish in this limit
\begin{equation}\label{eq:O-limit-for-propagation}
    \hat O_8 = [(\partial_-X_\parallel)^2 (\partial_+X_\parallel)^2]^2 \to 0 ~~~ as ~~~ X^{D-1}\to\infty ~~~\text{for~transmission}
\end{equation}
Suppose, without loss of generality, that this occcurs via $(J_\parallel^2)_{--}=(\partial_-X_\parallel)^2\to 0$ as $X^{D-1}\to\infty$.  Then the constraint from varying $S_{ws}$ with respect to the worldsheet metric, $\delta_h S=0$ would get no contribution from the $\lambda_W$ term.  So the constraints boil down to the usual conditions derived from $\delta_h S^{(0)}_{ws}=0$.  This includes the condition $T_{--}=0$, 
$(\partial_-X_\parallel)^2 = -(\partial_-X^{D-1})^2$.  

Thus $(\partial_-X^{D-1})^2$ would also have to also go to zero in this scenario. But this is impossible for the following reason.  A configuration with $\partial_- X^{D-1}=0$ would be purely right-moving for $X^{D-1}$, so it would be given by a mode expansion of the form $X^{D-1}=x^{D-1}+p^{D-1}\tau + w^{D-1} \sigma + \sum_n a_n e^{(\tau+\sigma)n/\ell_s}$ with $p^{D-1}=w^{D-1}$.    In order to propagate to $X_{D-1}\to\infty$, it must have momentum in the $X^{D-1}$ direction, $p^{D-1}>0$. But it cannot have winding in the $X^{D-1}$ direction since there is no periodicity in that direction (and anyway the string starts out as a closed loop in the bulk incoming region rather than stretching to infinity in $X^{D-1}$).  
% So in terms of the Fourier expansion $X^{D-1}=x^{D-1}+p^{D-1}\tau + w^{D-1} \sigma + oscillators$, $w=0$ and we need $p^{D-1}\ne 0$ to propagate to $X^{D-1}\to\infty$.
% There is no winding $w^{D-1}=0$ at finite energy since there is no topology.  
So it is not possible for $\partial_-X^{D-1}\to 0$.  As a result, the requirement \eqref{eq:O-limit-for-propagation} is not possible; the string reflects rather than oozing out to $X^{D-1}\to\infty$.  Since $\hat O$ does not vanish, contributions from the large $\lambda_W$ region are doubly exponentially suppressed.

It would be very interesting to complete the analysis of this type of theory similarly to Liouville and related examples.  We would like to know if the system will similarly complete to a CFT up to a renormalization of the $X^{D-1}$ dependence (including the value of $\kappa$).  In the full worldsheet theory, the path integration over 2d gravity formally produces some Weyl-invariant theory with zero net central charge.   
So far, our treatment confirms that if it completes to a theory whose matter CFT theory preserves the bulk region where the $\hat O$-Liouville effect is negligible, this system has a wall with total reflection.

\subsubsection{Vacuum $AdS_3$ cutoff wall}\label{sec:vac-AdS}

In order to move toward the spacetimes of interest, we may similarly consider the case of $AdS_3 \times S^3\times T^4$ with $k$ units of NS-NS flux \cite{Maldacena:2000hw, Maldacena:2000kv, Maldacena:2001km, Kutasov:1999xu, Giveon:1998ns}.  
In this case, we can start with the vacuum AdS metric
\begin{equation}\label{eq-adsvac-metric}
    ds^2 = d\rho^2 -\cosh^2\frac{\rho}{\ell} dt^2 + \sinh^2\frac{\rho}{\ell} d\phi^2 + internal
\end{equation}
along with $k$ units of background NS-NS field strength.  Following \cite{Maldacena:2000hw}, we write $u=\frac{1}{2}(t+\phi), v=\frac{1}{2}(t-\phi)$ in terms of which the currents are
$J_R^3=k(\partial_+ u + \cosh(2\rho)\partial_+v), J_R^\pm = k(\partial_+\rho \pm i \sinh(2\rho) \partial_+ v)e^{\mp i 2u}$ and similarly for the left movers with $u\leftrightarrow v$.  The AdS part of the right-moving stress-energy tensor is $T^{AdS}_{++}=\frac{1}{k}[\frac{1}{2}(J^+_R J^-_R+J^-_R J^+_R)-(J^3_R)^2]$.  Let us write
\begin{equation}\label{eq-DelSws-AdS}
    \Delta S_{ws} = -\lambda_W\int d^2\sigma  (\hat O_R+T_{++, internal})^2 (\hat O_L+T_{--, internal})^2\Phi_{m^2}(\rho, t,\phi)
\end{equation}
%For $\hat O_R$ here we can take e.g.
with
\begin{eqnarray}
    % \hat O_R & =& (J^+_R + J^-_R)^2-(J^+_R - J^-_R)^2 + (J^3_R)^2 \nonumber\\
    % &=& (\partial_+\rho\cos(2 u)+\sinh(2\rho)\partial_+ v\sin(2 u))^2+
    % (\partial_+\rho\sin(2 u)-\sinh(2\rho)\partial_+ v\cos(2 u))^2 \nonumber\\
    % & & ~~ + (\partial_+u + \cosh(2 \rho)\partial_+v)^2 \nonumber \\
    %\hat O_R & =& J^+_RJ^-_R + 
    \hat O_R = -(J^3_R)^2  = 
    % (\partial_+\rho)^2 + \sinh(2\rho)^2 (\partial_+ v)^2 + 
   -k^2(\partial_+u + \cosh(2 \rho)\partial_+v)^2  \nonumber \\
\end{eqnarray}
and similarly for $\hat O_L$; somewhat analogously to the flat example above, this is associated with the parallel directions to the wall ($u$ and $v$). 
Here, semiclassically $\Phi_{m^2}$ is the generalization of $\exp(\kappa X_3)$, i.e. a non-normalizable, time-independent solution of the bulk equation of motion for a scalar of mass squared $m^2$ corresponding to the dimension-8 operator in front.  
%Similarly one can solve for $\Phi$ here. 
At large $\rho \gg 1/\ell$ the equation simplifies to 
\begin{equation}
    2 \Phi' + \Phi'' = m^2 \Phi
\end{equation}
with dominant solution $\Phi\sim \exp\{(-1+\sqrt{1+m^2\ell^2})\rho/\ell\}\sim \exp(m \rho)$.

As in the examples above, we would like to check if this set up precludes propagation to large $\rho$.  The argument proceeds similarly. Given the exponential growth of $\Phi$ we would need a vanishing coefficient $\hat O = (\hat O_R+T_{++, internal})^2 (\hat O_L+T_{--, internal})^2 $.  Without loss of generality, we can consider this as resulting from the first factor vanishing; then the constraint that the total right-moving stress energy tensor $T_{++}^{AdS}+T_{++, internal}$ vanishes reverts to the condition $J^+ J^-+J^- J^+ = 0$, classically  $(\partial_+\rho)^2 + \sinh(2\rho)^2 (\partial_+ v)^2=0$.  
This in particular requires $\partial_+\rho=0$, which is in contradiction with the string propagating to large $\rho$ (similarly to $X^{D-1}$ in the case of the flat wall above).  
So at this level, we are finding a wall:  a radial cutoff in global $AdS_3$.    
We can adjust the position of the wall using the coefficient $\lambda_W$ of $\hat O$-Liouville term.   

Again here, it would be interesting to analyze this at a more detailed level, beyond establishing total reflection at the wall.  This example has a new feature compared to the flat wall case in \S\ref{sec:flat-flat}:  the bulk extrinsic curvature at a given $\rho$ does not vanish.  Nonetheless, the wall term in the worldsheet action is marginal at least at the semiclassical level, not sourcing any tadpoles at that order (in contrast to a generic deformation that would generate worldsheet beta functions indicating an immediate failure to solve target spacetime equations of motion).  Our application requires that the effective boundary in this system -- with appropriate higher-order renormalization of the function $\Phi$ -- remain at fixed radius, rather than developing nontrivial time dependence.  Given this, we will next comment on the implications for the effective low energy boundary condition.    

\subsection{The question of the effective boundary condition}\label{sec:OLT-effective-bc}

For holography including the deformation in \S\ref{sec:internal-equilibration}, we need to know whether or not the graviton and other low-energy bulk fields effectively has a fixed boundary condition at this type of Liouville wall (or if instead we integrate over boundary fields in the target spacetime, as in \cite{Horava:1996ma}).  We are also interested in the boundary conditions for heavy internal fields in the case of the compactification to AdS, as in our application in \S\ref{sec:internal-equilibration} the internal equilibration depends on their not having a Dirichlet boundary condition.\footnote{We thank Z. Yang for comments on this.}  

If the wall corresponded to a solution obtained by varying with respect to the fields everywhere, including the boundary, then there would be no net Brown-York energy, and no net electric NS-NS charge in the $AdS_3$ example.  The construction in \S\ref{sec:vac-AdS} exhibits nonzero extrinsic curvature and electric field at the wall, without an obvious source/sink of such fields introduced by the $\hat O$-Liouville term in the worldsheet action \eqref{eq-DelSws-AdS}.  
In particular, it is fundamental strings that source the NS-NS electric field, but the operator \eqref{eq-DelSws-AdS} is not a winding operator.  From this point of view, it appears that the wall is consistent with a fixed boundary condition for the low energy fields.

% In the case of Sin-Liouville theory, the deformation involves a condensation of momentum or winding modes (in T-dual descriptions). In that example, there is a duality to a cigar geometry which makes manifest that the boundary gravity is dynamical.   This is different from the present case, which seems consistent with a fixed boundary condition for the 3d graviton and NS-NS potential field.  

The internal degrees of freedom, including the moduli of the $S^3\times T^4$, are different in this regard:  they do not evolve radially.  This is consistent with a variety of effective boundary conditions, potentially enabling the internal fluctuations required for the deformation in \S\ref{sec:internal-equilibration}.  It is interesting to note that the aspects of gravitational Dirichlet walls which are UV sensitive -- involving arbitrarily high-momentum modes at the boundary \cite{Anderson:2006lqb}\cite{Witten:2018lgb} -- are softened by Liouville walls.

\section{Discussion and generalizations}\label{sec:discussion}

This paper investigated timelike boundaries in string/M theory, whose existence would
enable deformations between AdS and dS compactifications of string/M theory.  
The two parts of the paper are largely independent.

In the first part, we explained how given such boundaries, a naive obstruction to generalizing the deformation \cite{Coleman:2021nor} to higher dimensional quantum gravity is avoided in a very interesting way.  The matching between a radial AdS black hole boundary brought into the horizon and a dS static patch radial boundary occurs at very high temperature, leading to a thermal averaging in the extra dimensions which washes out the substantial internal differences.  So the major differences among internal string/M compactifications for dS and and for AdS are not an obstruction to this deformation procedure.  

In the second part of the paper, we explored a method to build bounding walls in string/M theoretic spacetimes.  We wrote down explicit models in flat and NS-NS vacuum $AdS_3$ spacetime using chiral currents and argued that these exhibit total reflection at the wall analogously to Liouville theory \cite{Teschner:2001rv}\cite{Seiberg:1990eb}.   

\subsection{Comments on black holes and dS in string-theoretic (A)dS}

We would like to similarly wall off black holes and the cosmic horizon in dS in order to upgrade the deformation between them \cite{Coleman:2021nor} to string theory as reviewed above in \S \ref{sec:internal-equilibration}.  An example of a pair of related AdS and dS models in type II string theory is given by the uplift of the D1-D5 system in \cite{Dong:2010pm}.  This involves systems with less symmetry and fewer chiral currents in the worldsheet sigma model compared to the above examples.  The fact that the wall thickness is much smaller than the curvature radius of the bulk may suggest that such generalizations exist, but this requires additional investigation.    

The general metric \eqref{eq:generalspace}, contains isometries along the putative boundary directions ($t$ and $\Omega$ translations).  The corresponding currents are not in general chiral, so the particular structure $\hat O\sim J_R ^4 J_L^4$ used above is not  available.  Perhaps one can use the chiral currents in the extended self-dual model \cite{Rocek:1991ps} whose gauging gives the original worldsheet sigma model upon integrating out the gauge field.  One could introduce a wall in the self-dual theory as above and study the effect of the gauging.  The viability of this depends on what the corresponding term $\hat O \Phi \sim J_R ^4 J_L^4 \Phi$ in the parent theory descends to in the gauged model.  Because $\hat O$ is irrelevant, the walled version of the self-dual model is non-Gaussian in the gauge field, a complication compared to the original analysis \cite{Rocek:1991ps}.  
A more direct approach may be to set up an approximate version of the above structure, taking $\hat O=(g^\parallel_{\mu\nu} \partial_+ X_\parallel^{\mu}\partial_+ X_\parallel ^\nu)^2(g^\parallel_{\rho\sigma} \partial_- X_\parallel^{\rho}\partial_- X_\parallel ^\sigma)^2$ with classical scaling dimension 8, and analyze its RG structure.

\subsection{Comments on the M theory case}

We would like to analyze this question also in M theory, with the goal of ultimately assessing the possibility of walls in the  the $dS_4$ models \cite{DeLuca:2021pej} obtained by uplift of the M2-brane theory.  

Let us start by considering the M-theory version of the flat wall in  \S\ref{sec:flat-flat}.  One approach would be to formulate an analogous wall in the BFSS matrix theory \cite{Banks:1996vh} in which the target spacetime is encoded in $U(N)$ matrices $X^M$. This would involve adding to the matrix theory Hamiltonian an operator of the form $Tr[\hat O_{M}\exp(\kappa X^{10})]$ whose effects die in the bulk.  In order to understand some of the constraints on this sort of construction (c.f. the constraint on $\kappa$ \eqref{eq:kappa-flat} in the worldsheet case), we may work with the matrix string theory of \cite{Dijkgraaf:1997vv} which interpolates between type IIA string theory and the eleven-dimensional limit of M theory.\footnote{We thank S. Shenker and L. Susskind for discussions on this case, and A. Frenkel for sharing other interesting ideas in this direction.}  

A Matrix string theory uplift of \eqref{eq:ws-def-flat-covar} would seem be
\begin{equation}
H = H_0 + Tr[\{(D_-X_\parallel-\frac{1}{4}\psi_\parallel\kappa \psi^D-1)^2(D_-X_\parallel-\frac{1}{4}\psi_\parallel\kappa \psi^D-1)^2\}^2 e^{\kappa X^{10}}]
\end{equation}
where $H_0$ is the maximally supersymmetric U(N) Yang Mills theory described in \cite{Dijkgraaf:1997vv} whose infrared limit is the $(R^8)^N/S_N$ theory of multiple free strings.  
In this framework, the constraint on $\kappa$ comes from requiring that the wall operator flow to a physical operator in the IR of this gauge theory, in each single-string sector arising from decoupled block in the matrix.  

\subsection{Summary}

The classification of timelike boundaries in string/M theory -- including in the case of positive cosmological constant -- is an important open problem.  If such walls exist in (A)dS backgrounds, descending at low energies to fixed boundary conditions for the bulk metric, it opens the possibility of holography for bounded patches that proceeds more similarly to AdS/CFT than either dS/CFT (with emergent time and fluctuating gravity) or dS/dS (with emergent radial spatial geometry, but fluctuating lower dimensional gravity).  

This idea is realized explicitly at the level of a solvable $T\bar T+\Lambda_2$ trajectory producing the emergent radial geometry of the $dS_3$ static patch and its refined entropy count \cite{Anninos:2020hfj, Coleman:2021nor, Shyam:2021ciy}\footnote{with ongoing work aimed at formulating local bulk non-gravitational matter \cite{WIP-bulk-matter}}.   Before the present work, however, that approach seemed doomed in the case of string/M theory, for which the internal compactification spaces descending to AdS and dS are quite different \cite{Flauger:2022hie}.  But as we explained in \S\ref{sec:internal-equilibration}, the resolution of this problem is already contained in the theory. Given the patch-bounding walls in string/M theory, we showed in \S\ref{sec:internal-equilibration} that the key step in the deformation from AdS/CFT to dS extends to the internal dimensions of string theory via the high-temperature averaging compelled at the (A)dS matching point in the trajectory where the boundary skirts the horizon.    

This question of fixed walls in sting/M theory is amenable to technical development.  In addition to the generalized Liouville theory  approach taken in \S\ref{sec:string-M-boundaries}, there may be other methods to formulate bounded systems with fixed boundary metric or extrinsic curvature (distinct from \cite{Horava:1996ma} which itself may also be interesting to generalize to additional backgrounds \cite{WITP-M-HW-walls}).\footnote{Perhaps the developments by Amr Ahmadain and A. Wall are particularly applicable \cite{Ahmadain:2022tew}.} We anticipate further progress on this well-motivated question.

\section*{Acknowledgements}

I would like to thank G. Torroba and S. Yang for comments on the manuscript. I am grateful to the authors of \cite{Coleman:2021nor} and the organizers and participants of the Corfu 2022 Workshop `Quantum features of a de Sitter Universe' for many useful discussions.  I am grateful to them and Shoaib Akhtar, Gauri Batra, G. Bruno De Luca, Alex Frenkel, Raghu Mahajan, Steve Shenker, Lenny Susskind, and Zhenbin Yang for useful remarks and/or ongoing collaboration on this topic.  This work is supported in part by a Simons Investigator award and National Science Foundation grant PHY-1720397.

\appendix

%\section{Conventions}\label{sec:notationconventions}

\bibliographystyle{JHEP}
\bibliography{refs.bib}
\end{document}